# Transport evidence of superlattice Dirac cones in graphene monolayer on twisted boron nitride substrate

Shimin Cao[1,2#], Mantang Chen[2#], Jiang Zeng[2§], Ning Ma[2], Runjie Zheng[2], Ya Feng[1], Shili Yan[1], Jing Liu[1], Kenji Watanabe[3], Takashi Taniguchi[3], X.C. Xie[1,2] and Jian-Hao Chen[1,2,4,5 †]

[1] *Beijing Academy of Quantum Information Sciences, Beijing 100193, China*
[2] *International Center for Quantum Materials, School of Physics, Peking University, Beijing 100871, China*
[3] *National Institute for Materials Science, Namiki 1-1, Tsukuba, Ibaraki 305-0044, Japan.*
[4] *Key Laboratory for the Physics and Chemistry of Nanodevices, Peking University, Beijing 100871, China*
[5] *Interdisciplinary Institute of Light-Element Quantum Materials and Research Center for Light-Element Advanced Materials, Peking University, Beijing 100871, China*

[§] *Current address: School of Physics and Electronics, Hunan University, Changsha 410082, China*
[†] *Corresponding author: Jian-Hao Chen <chenjianhao@pku.edu.cn>*
[#] *These authors contributed equally to this work.*

Strong band engineering in two-dimensional (2D) materials can be achieved by introducing moiré superlattices, leading to the emergence of various novel quantum phases with promising potential for future applications. Presented works to create moiré patterns have been focused on a twist embedded inside channel materials or between channel and substrate. However, the effects of a twist inside the substrate materials on the unaligned channel materials are much less explored. In this work, we report the realization of superlattice multi-Dirac cones with the coexistence of the main Dirac cone in a monolayer graphene (MLG) on a ~0.14° twisted double-layer boron nitride (tBN) substrate. Transport measurements reveal the emergence of three pairs of superlattice Dirac points around the pristine Dirac cone, featuring multiple metallic or insulating states surrounding the charge neutrality point (CNP). Displacement field tunable and electron-hole asymmetric Fermi velocities are indicated from temperature dependent measurements, along with the gapless dispersion of superlattice Dirac cones. The experimental observation of multiple Dirac cones in MLG/tBN heterostructure is supported by band structure calculations employing periodic moiré potential. Our results unveil the potential of using twisted substrate as a universal band engineering technique for 2D materials regardless of lattice matching and crystal orientations, which might pave the way for a new branch of twistronics.

The recently developed technique that enables the fabrication of van der Waals heterostructures with highly controlled orientations initiated a substantial surge in the research of moiré physics in these systems[1,2], which is coined as "twistronics"[3-9]. The first widely investigated system is twisted bilayer graphene (tBLG)[1,2,10-17]. Flat energy bands form near charge neutrality when the twist angle between two graphene monolayers is close to 1.08°, i.e. the so-called "first magic angle". The formation of flat bands plus the energy gap that isolates the flat bands from dispersive bands, lead to a clear manifestation of electron correlation and many-body ground states in this system. Phase diagram of tBLG near magic angle consists of a correlated insulating state surrounded by two superconducting domes[2,17], resembling high $T_c$ superconductors such as cuprates[18-20] and iron pnictides[21,22]. Beyond tBLG, plenty of other moiré structures are also studied both theoretically and experimentally, including homobilayers, heterobilayers or multilayers built out of hexagonal boron nitride (hBN)[23-25], transition metal dichalcogenides[26,27], 2D magnets[28], layered superconductors[29] and so on. Numerous exotic physical phenomena have been discovered, such as orbital magnetism, quantum anomalous Hall states and Chern insulators[10,11,17].

So far, most efforts in twistronics have been devoted to devices where the moiré potential is created between two layers of channel materials[1,2,14,26-37] or between the channel material and the substrate[38-41], which requires materials of the same symmetry and of highly similar lattice constants. In this paper, we study the modulation of the band structure of monolayer graphene by the moiré potential created by a twist of ~0.14° between two flakes of hBN substrates (tBN). The moiré periodic potential of such tBN substrate was created at the twisted interface[42], which is independent of the lattice constant, symmetry and alignment of the channel material. We found that the transport behavior of graphene 2nm away from the twisted interface is strongly modified, while the transport behavior of graphene 8nm away from the twisted interface is completely unaffected, which agrees with predicted exponential decay of the tBN moiré potential with distance[42]. Graphene on 2nm-tBN (G-2nm-tBN) shows multiple emergent insulating and metallic states surrounding the CNP, which are strongly tuned by the displacement field applied perpendicular to the heterostructure. Specifically, three pairs of prominent insulating states are observed, with each pair spread over both the electron doped and hole doped regions. Here "insulating states" mean that the temperature-dependent resistance is "insulating", i.e., resistance increases as temperature decreases. Further, we observed that the insulating states are the local maxima of the $R_{xx}$ vs. $n$ curves. In addition, the temperature dependent resistance of these insulating states does not fit to the thermal activation behavior, indicating the absence of a global gap. This is in agreement with density functional theory (DFT) calculations that predict the emergence of three pairs of resonant superlattice Dirac points with gapless spectra. On the other hand, the metallic states exhibit linear temperature dependent resistance, which agrees with expectations of electron-phonon scattering in a Dirac band[43,44]. Interestingly, such linear temperature dependence reveals Fermi velocities that are electrically tunable and electron-hole asymmetric. Thus, the twisted hBN substrate provides a significant band modulation for monolayer graphene, shedding light on the exploration of an alternative moiré generation method which could potentially be applied to a wide range of 2D materials.

Figure 1(a) illustrates the device structure in the experiment. A single crystal thin hBN flake with two regions of differing thicknesses (2nm and 8nm, see Supplementary Information I for details) was selected to fabricate the tBN substrates. The hBN sample was torn, twisted by a nominal 1° and then stacked to make tBN structures. Because of the slack in the gears of the mechanical rotator and potential relaxation at the distorted interface during the transfer process, the exact rotation angle is determined by transport measurement of the device to be ~0.14° (details in Supplementary Information III). During the process, two twisted interfaces were formed simultaneously: 1) the interface between 2nm hBN and ~0.14° twisted 2nm hBN (2nm-tBN); 2) the interface between 8nm hBN and ~0.14° twisted 8nm hBN (8nm-tBN). This tBN sample was then placed on top of a monolayer graphene without alignment using the conventional dry transfer technique, forming two device regions: graphene 2nm away from the twisted hBN interface (G-2nm-tBN) and graphene 8nm away from the twisted hBN interface (G-8nm-tBN), respectively. G-2nm-tBN and G-8nm-tBN were further encapsulated by two additional flakes of unaligned hBN (~50 nm in thickness). A graphite bottom gate and a Cr/Au top gate were also integrated into the heterostructure to form dual-gate devices. The optical micrograph of the device is shown in Figure 1(a). The blue shaded area indicates the G-2nm-tBN device while the brown shaded area corresponds to the G-8nm-tBN device.

Dual gate transport curves are obtained at 2 K for both G-2nm-tBN and G-8nm-tBN, as shown in Figure 1(b) and (c), respectively. In sample G-2nm-tBN, multiple resistance peaks appear at both sides of the main Dirac cone and their relative amplitudes vary with the bottom gate voltage, indicating electric displacement field tunability. For comparison, in sample G-8nm-tBN, only a single Dirac peak is observed in the response to gate voltages, identical to a normal hBN encapsulated graphene device. Since graphene has the same alignment to the tBN substrate for both G-2nm-tBN and G-8nm-tBN, normal transport behaviors in G-8nm-tBN rule out the possibility of unintentional alignment of graphene to the tBN, causing additional superlattice. The relatively large distance from graphene to moiré interface may account for the trivial resistance property in G-8nm-tBN sample, since the twist-induced periodic potential decays exponentially with distance[42]. From the $R_{xx}$-$V_{tg}$ curves of G-8nm-tBN we obtain a mobility exceeding $2 \times 10^5$ cm$^2$V$^{-1}$s$^{-1}$ at 2 K, indicating high quality of our samples. Thus the emergence of multiple resistance peaks in G-2nm-tBN shall be attributed to the effects from the moiré supercells of the tBN substrate, which is the major subject in later discussion.

It is worth noting that a similar work on graphene on stacking-engineered hBN substrate has been done recently[25]. Ref. [25] mainly focused on the ferroelectric response of parallel stacked bilayer hBN (P-BBN) up to a hysteresis level of $\Delta V_g / t_{BN} \approx 0.16$ V/nm, and they found that small angle (0.6°) twisted homobilayer hBN will only exhibit a much weaker ferroelectricity than that of P-BBN. No band engineering of the channel materials by the tBN moiré supercells is reported in Ref. [25].

In our experiment, no sign of ferroelectricity is observed for sweeping both top and bottom gate in two directions down to a hysteresis level of $\Delta V_g / t_{BN} \approx 0.0002$ V/nm (see Supplementary Information Figure S2). We attribute this to the odd-even layer number effect originated from the AA' stacking order of hBN flakes. In short, only odd-layer hBN will create strong ferroelectric potential at the interface when parallelly stacked to part of itself, while even-layer hBN does not has such effect. Very likely such ferroelectric response will diminish quickly with thicker hBN substrates. Detailed discussions on the absence of ferroelectricity in our devices can be found in Supplementary Information II. In the following we discuss band engineering of the channel materials by the tBN moiré supercells which was not experimentally discovered previously.

Figure 2(a) shows the carrier density $n$ and displacement field $D$ dependent resistance $R_{xx}$ of G-2nm-tBN. At $n = 0$, a resistance peak slightly tuned by $D$ could be identified (dotted line in Figure 2(a)). Accompanied by Hall measurement shown in Figure 2(b), this resistance peak is recognized as the main Dirac point. Around the main Dirac point, the $n$ and $D$ dependent $R_{xx}$ map exhibits a series of resistance peaks. The six most prominent peaks, marked by black arrows in Figure 2(a), are strongly dependent on $D$.

Figure 2(b) shows the $n$ and $D$ dependent Hall conductance $1/R_H$. Ripples appear in the Hall conductance, yet there is no global deviation from the straight lines, nor any additional sign reversal as observed in tBLG[14], twisted double bilayer graphene (tDBG)[45] or aligned graphene/hBN superlattice[46]. The Hall measurements imply that although tBN provides periodic potential, it does not fold the Dirac band completely and form global higher order Dirac points, nor does it open a band gap in the graphene channel. The moiré pattern of tBN substrate only induces subsidiary Dirac points together with an uneven density of states (DOS) in the graphene energy band, which would be further verified in the following temperature related analysis.

In order to better understand the features in $R_{xx}$ vs. $n$ and $D$ shown in Figure 2(a), we analyzed the change of resistance $\Delta R_{xx}$ vs. $n$ for fourteen different temperatures ranging from 0.4K to 32K and three different $D$ values as shown in Figure 2(c)-(e). Here $\Delta R_{xx}$ vs. $n$ curves are obtained by subtracting the smooth background of $R_{xx}(n, T=32\text{K})$ from the lower temperature curves $R_{xx}(n, T<32\text{K})$. In particular, Figure 2(c)-(e) show $\Delta R_{xx}(n)$ curves for displacement field $D = +0.18$ V/nm, 0 V/nm and -0.18 V/nm, respectively. Similar to Figure 2(a), the main Dirac point is marked by black dotted lines and the six prominent resistance peaks by black arrows in Figure 2(c)-(e). We shall discuss two of these six resistance peaks in the main text (marked by black triangles in Figure 2(c)-(e)) while the other four resistance peaks are discussed in Supplementary Information IV. One important observation is that these resistance peaks are all *insulating*, i.e., $\Delta R_{xx}$ increases as the sample temperature deceases. Interestingly, at the immediate vicinity of the main Dirac point, two prominent resistance "valleys" are also visible. One valley is located at the hole side and the other at the electron side, which are marked by blue triangles in Figure 2(c)-(e). Such resistance valleys are *metallic*, i.e., $\Delta R_{xx}$ decreases as the sample temperature deceases.

Next, we carefully study the two insulating states and the two metallic states, marked by black triangles and blue triangles in Figure 2(c)-(e), respectively. Figure 3(a)&(b) show the Arrhenius plots of the two insulating states, displaying $\log(R_{xx}/R_0)$ vs. inverse temperature. Here $R_{xx}$ is the longitudinal resistance of the two resistance peaks, and $R_0$ is the resistance at $T = 32$ K. As can be seen in Figure 3(a)&(b), the temperature dependence of the insulating states does not fit to the thermal activation behavior (i.e., $R_{xx} \sim \exp(-E_g/2k_B T)$), therefore the graphene spectrum under tBN is indeed gapless even at the superlattice Dirac points, consistent with Hall measurement results. Figure 3(c)&(d), on the other hand, show $R_{xx}$ vs. $T$ curves of the two metallic states at the hole side and electron side of the main Dirac point. It can be seen that for the two metallic states, $R_{xx}$ is linear with $T$, which is a common behavior in similar graphene moiré systems[32,44,45]. Such a linear $T$ dependence is a consequence of acoustic phonon scattering of charge carriers[44,47]. Resistivity due to quasi-elastic scattering of acoustic phonons in a graphitic system can be expressed as:

$$\rho = \frac{\pi F D_A^2}{g e^2 \hbar \rho_m v_F^2 v_{ph}^2} k_B T \qquad \text{Eq.(1)}$$

where $F$ is the form factor accounting for the scattering matrix elements of differing electron-phonon processes. In the case of monolayer graphene, $F = 1$. $D_A = 18\text{eV}$ is the deformation potential describing the electron-phonon scattering strength. $g = 4$ is the degree of degeneracy of spin and valley for monolayer graphene and $\rho_m = 7.6 \times 10^{-7}$ kg/m² is the mass density of graphene. $v_F$ and $v_{ph} = 2 \times 10^4$ m/s are Fermi velocity and phonon group velocity in graphene, respectively[48]. Since $D_A$ and $v_{ph}$ reflect intrinsic properties of the atomic lattice of graphene, it is reasonable to assume both parameters are invariant with the applied displacement field $D$ or carrier density $n$. Therefore, the slope of the linear $R_{xx}$-$T$ relationship $S = dR/dT \propto 1/v_F^2$ offers us an estimation of the Fermi velocity of this system. As depicted in Figure 3(e), both of such slopes at the hole side ($S_H$) and the electron side ($S_E$) clearly drop with the increment in the absolute strength of displacement field $|D|$ (as shown by the dashed arrows in Figure 3(e)), indicating an enhanced $v_F$ for non-zero $|D|$. Meanwhile, the ratio of $S_H$ to $S_E$ shows a monotonic dependence on $D$ (Figure 3(f)). Figure 3(g) shows the extracted Fermi velocities for both hole doping ($v_{F,H}$) and electron doping ($v_{F,E}$) vs. $D$, showing $v_{F,H} > v_{F,E}$ ($v_{F,H} < v_{F,E}$) for $D < 0$ ($D > 0$), respectively. As $D$ varies from -0.35 V/nm to 0.35 V/nm, the ratio of $v_{F,E}$ to $v_{F,H}$ increases monotonically (Figure 3(h)), revealing an electrically tunable electron-hole asymmetry in the tBN modulated monolayer graphene.

To understand the formation mechanism of the emergent insulating states and metallic states which are highly dependent on $D$, we employed band calculations of the MLG/tBN heterostructure based on DFT (see Supplementary Information III for more details). It has been shown that twisted double-layer hBN forms a moiré pattern, which acts as periodic potential and can be used to engineer the electronic properties of the channel materials at proximity to it[25,42,49]. Here we consider the model Hamiltonian[38,50]:

$$H = t \sum_{<i,j>} a_i a_j^\dagger + V \sum_{\alpha,i} \cos(\boldsymbol{G}_\alpha \boldsymbol{x}_i) n_i \qquad \text{Eq. (2)}$$

where $t$ is the magnitude of the nearest neighbor hopping on the graphene honeycomb lattice, $V$ is the magnitude of the periodic potential and the reciprocal vectors $\boldsymbol{G}_\alpha$ ($\alpha$=1,2,3) are determined by the moiré pattern of the twisted hBN interface. $|\boldsymbol{G}_\alpha| = 4\pi/\sqrt{3}\lambda$, where $\lambda$ is the wavelength of the moiré superlattice determined by the twist angle between the two layers of hBN. The three wavevectors $\boldsymbol{G}_{1,2,3}$ are 120° rotated with respect to each other. A prototype 81×81 supercell of hBN is constructed with periodic length $\lambda \sim 20$nm in our calculation, which is also the spatial period of the superlattice potential. Based on our *ab-initio* calculation, we adopted the typical values of $t = 2.7$ eV and $V = 30$ meV and obtained the low energy band structures of tBN supported MLG as shown in Figure 4. Indeed, in addition to the pristine Dirac cone (D0), multiple superlattice Dirac cones emerge in the electronic band of a graphene monolayer on the tBN substrate (labeled as D1, D2, D3), and there is no global band gap. From a perturbative viewpoint, the mentioned three pairs of prominent superlattice Dirac cones, together with dips in the DOS profile, arise from resonance effects[38,51]. These subsidiary Dirac cones are in accord with the experimental observation of three pairs of insulating resistance peaks as shown in Figure 2(a). Note that the above calculation employed a simplified model with a relatively small moiré supercell, corresponding to a twist angle of ~0.7°. This simplified model qualitatively agrees to the experiment results, indicating that the main physics is captured. Quantitative agreement to experiment is expected when the computational moiré supercell is ~25 times bigger (see Supplementary Information III), corresponding to a twist angle of ~0.14°, which is computationally too costly. Combining the experimental results and theoretical calculations, we conclude that the moiré potential from tBN substrate could effectively induce multiple superlattice Dirac cones near the main Dirac point of monolayer graphene. Furthermore, such moiré potential could lead to electric displacement field tunable electron-hole asymmetry in graphene, as revealed by the adjustable Fermi velocities in Figure 3(e)-(h).

In conclusion, we have fabricated the heterostructure consisting of twisted hBN substrate (twist angle ~0.14°) interfaced with unaligned monolayer graphene, and we systematically studied the transport behavior of the sample. It is experimentally ensured that such an unaligned graphene-tBN by itself does not result in any distinctive transport properties in graphene, while the twisted hBN interface creates moiré potential that would strongly modify the graphene band structure. When graphene is 2nm away from the tBN interface, gapless superlattice Dirac cones surrounding the pristine Dirac point are observed in transport measurements, which is corroborated by DFT calculations. Fermi velocities near the main Dirac point are highly tunable by the applied displacement field and are electron-hole asymmetric. Our results unveil the potential of twisted hBN substrate to provide a universal moiré modulation for various types of 2D materials, which offers an alternative approach in search of exotic physical phenomena in twistronics. Compared to the widely studied moiré systems either generated from "twisted channels" [1,2,14,26-37] or from "twisted channel and substrate" [38-41], the "twisted substrate" method is free of the lattice matching and symmetry matching requirements.

**Acknowledgement**

This project has been supported by the National Basic Research Program of China (Grant Nos. 2019YFA0308402, 2018YFA0305604, 2015CB921102, 2015CB921100), the National Natural Science Foundation of China (NSFC Grant Nos. 11934001, 92265106, 11774010, 11921005, 11534001), Beijing Municipal Natural Science Foundation (Grant No. JQ20002), the Strategic Priority Research Program of Chinese Academy of Sciences (Grant No. XDB28000000).

**Contributions**

J.-H.C. and S.C. conceived the experiment; S.C. fabricated the device and performed most of the measurement; N.M. and M.C. aided in device fabrication; M.C. and S.C. analyzed the data; R.Z., J.L., Y.Z. and S.Y. aided in transport measurement; J.Z. and X.C.X. provided theoretical analysis; K.W. and T.T. grew high quality boron nitride bulk crystals; S.C., J.Z., M.C. and J.-H.C. wrote the manuscript and all authors commented and modified the manuscript.

**References**

[1] Y. Cao *et al.*, Nature **556**, 80 (2018).
[2] Y. Cao, V. Fatemi, S. Fang, K. Watanabe, T. Taniguchi, E. Kaxiras, and P. Jarillo-Herrero, Nature **556**, 43 (2018).
[3] E. J. Mele, Physical Review B **81**, 161405 (2010).
[4] E. J. Mele, Physical Review B **84**, 235439 (2011).
[5] M. Kindermann and E. J. Mele, Physical Review B **84**, 161406 (2011).
[6] A. Pal and E. J. Mele, Physical Review B **87**, 205444 (2013).
[7] R. Bistritzer and A. H. MacDonald, Physical Review B **81**, 245412 (2010).

[8] R. Bistritzer and A. H. MacDonald, Proceedings of the National Academy of Sciences **108**, 12233 (2011).
[9] R. Bistritzer and A. H. MacDonald, Physical Review B **84**, 035440 (2011).
[10] A. L. Sharpe, E. J. Fox, A. W. Barnard, J. Finney, K. Watanabe, T. Taniguchi, M. A. Kastner, and D. Goldhaber-Gordon, Science **365**, 605 (2019).
[11] M. Serlin, C. L. Tschirhart, H. Polshyn, Y. Zhang, J. Zhu, K. Watanabe, T. Taniguchi, L. Balents, and A. F. Young, Science **367**, 900 (2020).
[12] X. Lu *et al.*, Nature **574**, 653 (2019).
[13] J. R. Wallbank *et al.*, Nature Physics **15**, 32 (2019).
[14] Y. Saito, J. Y. Ge, K. Watanabe, T. Taniguchi, and A. F. Young, Nature Physics **16**, 926 (2020).
[15] G. Li, A. Luican, J. M. B. Lopes dos Santos, A. H. Castro Neto, A. Reina, J. Kong, and E. Y. Andrei, Nature Physics **6**, 109 (2010).
[16] M. Yankowitz, S. Chen, H. Polshyn, Y. Zhang, K. Watanabe, T. Taniguchi, D. Graf, A. F. Young, and C. R. Dean, Science **363**, 1059 (2019).
[17] P. Stepanov *et al.*, Nature **583**, 375 (2020).
[18] P. A. Lee, N. Nagaosa, and X.-G. Wen, Reviews of Modern Physics **78**, 17 (2006).
[19] N. P. Armitage, P. Fournier, and R. L. Greene, Reviews of Modern Physics **82**, 2421 (2010).
[20] B. Keimer, S. A. Kivelson, M. R. Norman, S. Uchida, and J. Zaanen, Nature **518**, 179 (2015).
[21] G. R. Stewart, Reviews of Modern Physics **83**, 1589 (2011).
[22] Q. Si, R. Yu, and E. Abrahams, Nature Reviews Materials **1**, 16017 (2016).
[23] C. R. Woods *et al.*, Nat Commun **12**, 347 (2021).
[24] M. Vizner Stern *et al.*, Science (2021).
[25] K. Yasuda, X. Wang, K. Watanabe, T. Taniguchi, and P. Jarillo-Herrero, Science (2021).
[26] L. Zhang *et al.*, Nature Communications **11** (2020).
[27] L. P. McDonnell, J. J. S. Viner, D. A. Ruiz-Tijerina, P. Rivera, X. Xu, V. I. Fal'ko, and D. C. Smith, 2D Materials **8**, 035009 (2021).
[28] H. Xie *et al.*, Nature Physics **18**, 30 (2022).
[29] L. S. Farrar, A. Nevill, Z. J. Lim, G. Balakrishnan, S. Dale, and S. J. Bending, Nano Letters **21**, 6725 (2021).
[30] G. W. Burg, J. Zhu, T. Taniguchi, K. Watanabe, A. H. MacDonald, and E. Tutuc, Physical Review Letters **123**, 197702 (2019).
[31] C. Shen *et al.*, Nature Physics **16**, 520 (2020).
[32] Y. Cao, D. Rodan-Legrain, O. Rubies-Bigorda, J. M. Park, K. Watanabe, T. Taniguchi, and P. Jarillo-Herrero, Nature **583**, 215 (2020).
[33] X. Liu *et al.*, Nature **583**, 221 (2020).
[34] Z. Hao, A. M. Zimmerman, P. Ledwith, E. Khalaf, D. H. Najafabadi, K. Watanabe, T. Taniguchi, A. Vishwanath, and P. Kim, Science **371**, 1133 (2021).
[35] J. M. Park, Y. Cao, K. Watanabe, T. Taniguchi, and P. Jarillo-Herrero, Nature **590**, 249 (2021).
[36] E. C. Regan *et al.*, Nature **579**, 359 (2020).
[37] Y. Tang *et al.*, Nature **579**, 353 (2020).
[38] M. Yankowitz, J. Xue, D. Cormode, J. D. Sanchez-Yamagishi, K. Watanabe, T. Taniguchi, P. Jarillo-Herrero, P. Jacquod, and B. J. LeRoy, Nature Physics **8**, 382 (2012).
[39] L. A. Ponomarenko *et al.*, Nature **497**, 594 (2013).
[40] Z. Zheng *et al.*, Nature **588**, 71 (2020).

[41] G. Chen *et al.*, Nature **579**, 56 (2020).

[42] P. Zhao, C. Xiao, and W. Yao, npj 2D Materials and Applications **5**, 1 (2021).

[43] E. H. Hwang and S. Das Sarma, Physical Review B **77** (2008).

[44] H. Polshyn, M. Yankowitz, S. Chen, Y. Zhang, K. Watanabe, T. Taniguchi, C. R. Dean, and A. F. Young, Nature Physics **15**, 1011 (2019).

[45] M. He, Y. Li, J. Cai, Y. Liu, K. Watanabe, T. Taniguchi, X. Xu, and M. Yankowitz, Nature Physics **17**, 26 (2020).

[46] X. Lu *et al.*, Physical Review B **102**, 045409 (2020).

[47] F. Wu, E. Hwang, and S. Das Sarma, Phys. Rev. B **99**, 165112 (2019).

[48] J.-H. Chen, C. Jang, S. Xiao, M. Ishigami, and M. S. Fuhrer, Nature Nanotechnology **3**, 206 (2008).

[49] X.-J. Zhao, Y. Yang, D.-B. Zhang, and S.-H. Wei, Physical review letters **124**, 086401 (2020).

[50] C.-H. Park, L. Yang, Y.-W. Son, M. L. Cohen, and S. G. Louie, Nature Physics **4**, 213 (2008).

[51] C.-H. Park, L. Yang, Y.-W. Son, M. L. Cohen, and S. G. Louie, Phys. Rev. Lett. **101**, 126804 (2008).

# Transport evidence of superlattice Dirac cones in graphene monolayer on twisted boron nitride substrate

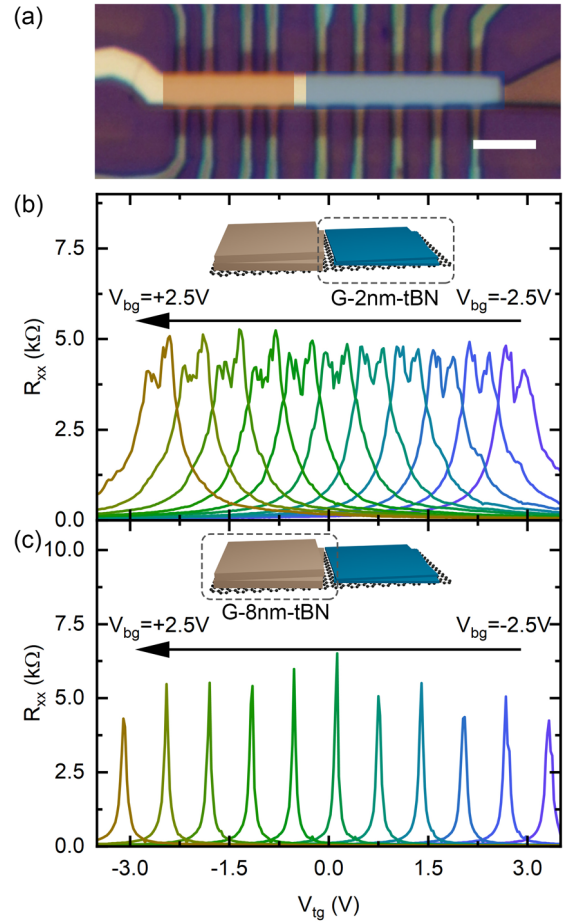

**Figure 1. Longitudinal resistance of monolayer graphene on twisted boron nitride (tBN). (a)** Optical micrograph of the MLG/tBN heterostructure device with twist angle ~ 0.14°. Blue shaded area indicates 2nm-thick part of hBN, and brown shaded area indicates 8nm-thick part of hBN. Scale bar is 5μm. **(b-c)** Gate voltage dependent $R_{xx}$ of (b) G-2nm-tBN and (c) G-8nm-tBN at $T$ = 2 K. Curves from right to left correspond to bottom gate voltage varying from -2.5 V to +2.5 V, in 0.5 V steps. Insets: schematic illustration of the tBN decorated monolayer graphene heterostructure.

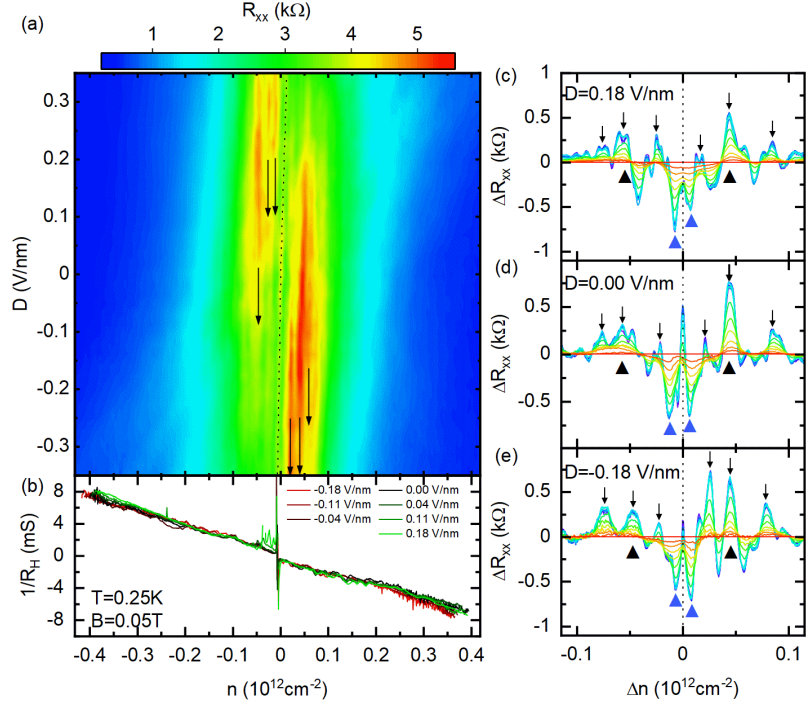

**Figure 2. Carrier density and displacement field dependent transport of the G-2nm-tBN sample. (a)** Longitudinal resistance at $T$=250mK as a function of displacement field $D = \epsilon_{BN}(V_{tg}/d_{tg} - V_{bg}/d_{bg})/2$ and carrier density $n = \epsilon_0\epsilon_{BN}(V_{tg}/d_{tg} + V_{bg}/d_{bg})/e$, where $\epsilon_{BN} = 3.5$ is the relative permittivity of hBN, $V_{tg}$ ($V_{bg}$) and $d_{tg}$ ($d_{bg}$) are gate voltage and thickness of the encapsulating hBN on top (bottom), respectively. The dotted line denotes the main Dirac point, and black arrows mark emergent insulating states around it. **(b)** Carrier density and displacement field dependence of the Hall conductance. **(c-e)** The carrier density dependence of $\Delta R_{xx} = R_{xx}(T) - R_{xx}(T = 32\text{ K})$ at (c) $D = +0.18$ V/nm, (d) $D = 0$ V/nm and (e) $D = -0.18$ V/nm. Different curves correspond to temperature from 0.4K (purple) to 32K (red). $\Delta n$ is the nominal carrier density with respect to the charge neutrality point. Dotted lines denote the main Dirac point, while black arrows indicate the surrounding insulating states. Black and blue triangles mark the insulating and metallic states discussed in the main text, respectively.

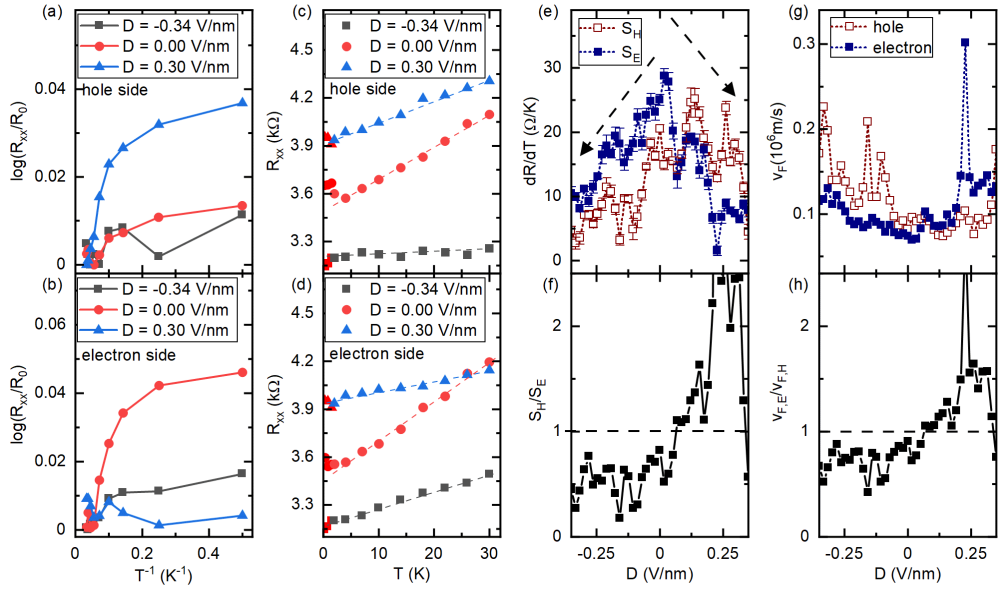

**Figure 3. Temperature dependent behavior of the insulating and metallic states of the G-2nm-tBN sample. (a-b)** Arrhenius plot of the insulating states at the hole side (a) and at the electron side (b), marked by black triangles in Figure 2(c)-(e). **(c-d)** Linear $R$-$T$ relationship of the metallic states at the hole side (c) and at the electron side (d), marked by blue triangles in Figure 2(c)-(e). Dashed lines are guide for the eye. **(e)** $dR/dT$ as a function of displacement field $D$ for the metallic states at the hole side ($S_H$) and at the electron side ($S_E$) and **(f)** their ratio. Dashed arrows in Figure 2(e) imply the trend of a dropping slope, or an increasing $v_F$, with increasing $|D|$. **(g)** Experimentally extracted $v_F$ at the electron side and at the hole side and **(h)** their ratio.

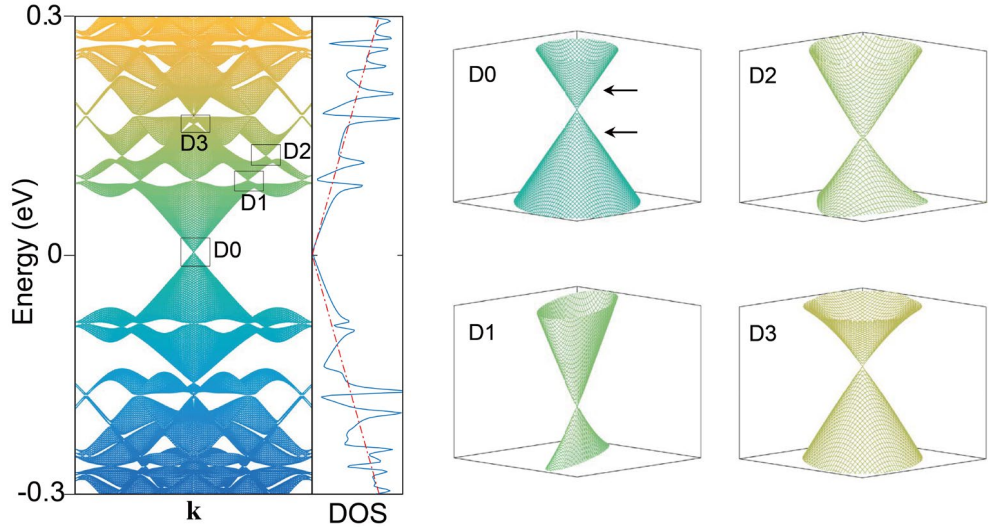

**Figure 4. Multiple Dirac cones in monolayer graphene on tBN substrate.** The left panel shows calculated band dispersion and density of states of monolayer graphene on tBN for an 81×81 supercell and with $V = 30$ meV. The pristine Dirac cone of graphene and three superlattice Dirac cones above it are labeled as D0, D1, D2 and D3 respectively, with their zoom-in spectra shown in the right panels. Black arrows pointing at D0 cone indicate the energy levels of the two metallic states discussed in the main text.